\documentclass{PoS}
\usepackage[utf8]{inputenc} 
\usepackage[T1]{fontenc} 
\usepackage{amssymb,amsfonts,amsmath,amsthm} 
\usepackage{graphicx} 
\usepackage{textcomp}
\usepackage{dsfont}
\usepackage{slashed}	

\newcommand{\lvac}{\langle\; {\underset{\raise0.3em\hbox{$\smash{\scriptscriptstyle\thicksim}$}}{0} }\;|}
\newcommand{\rvac}{|\; {\underset{\raise0.3em\hbox{$\smash{\scriptscriptstyle\thicksim}$}}{0} }\; \rangle}
\newcommand{\eins}{\mathds{1}}	
\newcommand{\nc}{N_\text{c}}

\title{{\footnotesize\mdseries TTP10-39\\[-2ex] DO-TH 10/15}\\[1.5ex] Charm mixing in the Standard Model}

\ShortTitle{Charm mixing in the Standard Model}

\author{\speaker{Markus Bobrowski}\thanks{This work has been supported by the DFG Research Unit SFB/TR9 and by grants of the German National Academic Foundation (Studienstiftung des deutschen Volkes), the State of Bavaria, and the German National Academic Exchange Service (DAAD).}\\%
        Institute for Theoretical Particle Physics, Karlsruhe Insitute of Technology\\
        76128 Karlsruhe, Germany\\
        E-mail: \email{markus.bobrowski@kit.edu}}

\author{Alexander Lenz\\
        Theoretical Physics T3, TU Dortmund\\
        44221 Dortmund, Germany\\
        E-mail: \email{alexander.lenz@tu-dortmund.de}}

\abstract{In this talk we report on a study on the mixing of neutral charmed mesons and argue that, at the present stage, a CP violating weak phase of the order of some per mille can not be excluded in the Standard Model. It is shown how some, seemingly reasonable, simplifying assumptions about CKM couplings lead to the wrong conclusion that CP violation of this amount is an unambiguous indication of new physics. \newline
The presented results rely on a recent short-distance analysis of the $\Delta C=2$ transition, which confirms the expectation that the dominant contribution is due to effects of flavour symmetry breaking appearing in higher orders of the Heavy Quark Expansion. We investigate meson-antimeson transitions with an intermediate state coupling to the meson's sea quark background, present in dimension 10 and 12, using a factorisation approach to simplify the operator basis. On account of a lifting of GIM suppression by one power of $m_s/m_c$, the contribution to $y=\Delta\Gamma/2\Gamma$ is found to exceed that of the formally leading dimension six by a factor close to ten.}

\FullConference{35th International Conference of High Energy Physics\\
                 July 22-28, 2010\\
                 Paris, France}

\begin{document}

\section{Introduction}

Flavour oscillations of neutral mesons arise in presence of non-zero mass and width differences $\Delta M$ and $\Delta \Gamma$ between the long- and short-lived mass eigenstate components. First evidence for $D^0$ oscillations was reported in 2007 by \emph{Belle} and \textsc{BaBar}, and later confirmed by CDF \cite{exp}. HFAG quotes the current world average for the $D^0$ mixing rates to be $x\equiv\Delta M/\Gamma=\left( {0.59 \pm 0.20} \right)\%$ and $y\equiv\Delta\Gamma/2\Gamma=\left( {0.80 \pm 0.13} \right)\%$ \cite{Barberio:2008fa}. 
On the theory side, two approaches are known to predict these quantities: in the inclusive approach, the $\Delta C=2$ transition  amplitude is determined at quark-level in the full Standard Model \cite{papersCds,inclusive}. The calculation relies on the Heavy Quark Expansion (HQE), an expansion of the effective Hamiltonian into a series of local operators of increasing dimension. 
An alternative possibility is to obtain the width difference from an exclusive sum over hadronic intermediate states \cite{exclusive}, requiring the knowledge of many decay amplitudes and strong phases to a high precision. %The available works are limited to on an estimation of $\text{SU}(3)_\text{F}$ breaking from phase space, assuming the absence of cancellations between $\text{SU}(3)_\text{F}$  breaking from phase space and $\text{SU}(3)_\text{F}$  matrix elements and between different $\text{SU}(3)_\text{F}$ multipletts. 
At present, none of these approaches leads to completely satisfying results.

\section{Dimension six and seven}
Diagonalisation of the $2\times 2$ mixing Hamiltonian $\hat H =\hat M+ \text{i}\, \hat\Gamma/2$  relates mass and width difference to the off-diagonal elements $M_{12}$ and $\Gamma_{12}$. This study reports on a calulation of $\Gamma_{12}$ in an HQE framework, \textit{i.e.} as an expansion into a series of operators ${\mathbf{Q}}_D$ of mass dimension $D$:
\begin{equation}
 \Gamma_{12}  = \frac{1}
{{2M_{D^0} }}\;{\rm Im} \;{\rm i}\int {{\rm{d}}^4 x\,\left\langle\bar D^0 \right| \,{{\rm T}\;\mathcal{H}\left( x \right)\mathcal{H}\left( 0 \right)} \,\left| D^0 \right\rangle  = \sum\limits_{D = 0}^\infty  {\left( {\frac{\Lambda }
{{m_{ c} }}} \right)^D \;{\mathbf{G}}_D \left\langle {{\mathbf{Q}}_D } \right\rangle } } .
\end{equation}
First contributions to $\Gamma_{12}$ appear for $D=6$, associated with the absorptive part of the $\Delta C=2$ box diagram. Prior to this work, the Wilson coefficients $\mathbf{G}_D$ have been known up to $D=8$ and at next-to-leading order QCD. The GIM mechanism, an effect of CKM hierarchy and residual $\text{SU}(3)_\text{F}$ quark flavour symmetry, requires these contributions to be very small in the Standard Model: expanding with respect to the CKM structure and using the unitarity of the CKM matrix, $\Gamma_{12}$ can be written as
\begin{equation}\label{10081701}
\Gamma _{12}  =  - \lambda _s^2 \;\left( {\Gamma _{12}^{ss}  - 2\,\Gamma _{12}^{sd}  + \Gamma _{12}^{dd} } \right) + 2\,\lambda _s \lambda _b \;\left( {\Gamma _{12}^{sd}  - \Gamma _{12}^{dd} } \right) - \lambda _b^2 \;\Gamma _{12}^{dd} ,
\end{equation}
where $\lambda_q=V_{cq}^*V_{uq}$. The CKM couplings $\lambda_s=\mathcal{O}(\lambda)$ and $\lambda_b=\mathcal{O}(\lambda^5)$ induce a  hierarchy in powers of  the Wolfenstein parameter $\lambda \simeq 0.2255$. In the limit of exact $\text{SU}(3)_\text{F}$ symmetry the linear combinations  in brackets cancel to zero. $\text{SU}(3)_\text{F}$ breaking effects enter as terms proportional to powers of $\overline  z = \overline  m_s^2 (\overline  m_c)/\overline  m_c^2 (\overline  m_c)$. Residual flavour symmetry gives rise to cancellations down to terms of order $\overline  z^2$ in the CKM-leading, and of order $\overline  z$, respectively, in the CKM-subleading contribution, \textit{i.e.} 
\begin{equation}\label{10081801} 
	\begin{gathered}
	  \Gamma _{12}  =  - (1.15\;\overline  z^2-59.7 \;\overline  z^3 )\;\lambda _s^2   - 5.5\;\lambda _s \lambda _b \;\overline  z - 1.96\;\lambda _b^2  \; \;   \simeq \;\;\lambda ^{9.0}  \;+\; \lambda ^{8.0}  \;+\;\lambda ^{9.5}.  
	\end{gathered} 
\end{equation}
Numerically, we find $\Gamma_{12} = -(0.20-0.16\,\text{i})\cdot 10^{-5}$, and accordingly $y\leq |\Gamma_{12}| \cdot \tau_D \simeq 10^{-6}$ \cite{Bobrowski:2010xg}, in agreement with previous studies. If the experimental average for $y$ remains at its current central value, this is orders of magnitude too small. We remark that commonly   the small imaginary parts of $\lambda_d$ and $\lambda_s$ are neglected, which is equivalent to setting $\lambda_b/ \lambda_s\simeq 0$ in (\ref{10081701}). $\Gamma_{12}$ then is found to be real with high accuracy. Yet actually  $\text{SU}(3)_\text{F}$  symmetry is efficient enough to make the second term in (\ref{10081701}) even exceed the first one, such that the approximation $\lambda_b \ll \lambda_s$ certainly is not justified. Keeping all CKM factors exactly, $\operatorname{arg}\lambda_b =1.17$ introduces an order one phase to $\Gamma_{12}$. We therefore think that strong claims,  CP violation of the order of 1\textperthousand~was an unambiguous sign of new physics, should be met with some caution. 
As to an understanding of the remarkable deviation of the $D=6,\,7$ quark-level result from experiment, four main lines of argumentation are common: $^{(1)}\,$One could suspect the malfunction of heavy quark methods at the charm threshold, where QCD may become unmanageable and the inverse quark mass may no longer be a suitable expansion parameter. 
To investigate  the behaviour of the perturbative expansion in QCD and $1/m_c$, we calculated the $\mathcal{O}(\alpha_s)$ and $D=7$ corrections to the $\Gamma_{12}^{ab}$ and found them to be below $\sim25\%$. Although we certainly must not expect a precision prediction of $D^0$ mixing rates, this does not seem to us as as indication for a breakdown of the expansion. $^{(2)}\,$Valence quark dy\-namics may cease to offer a reliable description for the hadron-level transition and non-per\-tur\-ba\-tive long-distance dynamics, violating quark hadron duality, may become important. The dominant con\-tri\-bu\-tions to the $\Delta C=2$ transition in this case would not be captured by the HQE approach. $^{(3)}\,$There are reasons to expect  $\text{SU}(3)_\text{F}$-breaking effects in higher orders of the HQE, which could dras\-ti\-cal\-ly enhance $\Gamma _{12}^{ss}  - 2\,\Gamma _{12}^{sd}  + \Gamma _{12}^{dd}$ and $\Gamma _{12}^{sd}-\Gamma _{12}^{dd}$. $^{(4)}\,$And finally: also new physics could be responsible, \textit{e.g.} by violating the $3\times 3 $ unitarity of the CKM matrix \cite{Bobrowski:2009ng} or by intruducing right-handed charged currents. 

A more definite statement about the reliability of heavy quark methods  could be provided by a future calculation of charmed meson decay widths; they receive the leading contribution from the spectator model charm decay $\Gamma_0(c)$ in $D=3$, where GIM suppression is absent. %---making higher orders and new physics contributions safely negligible. 
For the time being, a first estimate can be obtained from a comparison of lifetime ratios $\tau(D^+)/\tau(D^0)\simeq 2.5$ and $\tau(D_s^+)/\tau(D^0)\simeq 1.2$ to experiment: writing $\Gamma =\Gamma_0(c)(1+\delta)$ and neglecting the small difference in phase space between the decays of $D^+$ and $D^+_s$, we extract that the  leading order HQE   is off by  $\delta(D^0) =+17\%$, $\delta(D^+) =-53\%$ and $\delta(D_s^+) =-3\%$, respectively. Albeit in no way compulsive, this supports the expectation that the HQE should at least reproduce the correct order-of-magnitude. Note that this estimation is largely free of hadronic uncertainties, which cancel with $\Gamma_0(c)$.

\section{Flavour symmetry breaking in higher dimensions} 
\begin{figure}
	\centering
		\includegraphics{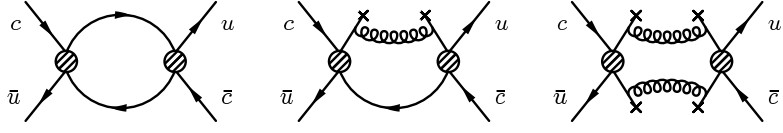}
	\caption{Contributions to the $\Delta C=2$ transition formally appearing in dimension 6, 9 and 12 in the HQE. Crosses indicate intermediate states coupling to the diquark condensate.}
	\vspace{-1.5ex}
	\label{fig:cdsLandscape}
\end{figure}

Formally, the smallness of the $\Delta C=2$ box diagrams is due to the fact that one mass insertion per internal fermion line is required to break the $\text{SU}(3)_\text{F}$  flavour interference in (\ref{10081701}), and a second one to compensate the chirality flip---leading to the double GIM suppression present in (\ref{10081801}). It should, in accordance, be possible to lift one order of GIM suppression by cutting one of the internal fermion lines, allowing the loose ends to couple to the meson's sea quark and gluon background. Diagrams of this kind appear in dimension $D\geq 9$ and $D\geq 12$, respectively, and commonly have been expected to be the dominant contributions within the HQE \cite{papersCds}. 

This work considers diagram topologies with one intermediate state coupling to the low-energy background (\textit{cf}. Fig. \ref{fig:cdsLandscape}),  contributing at $D\geq 9$. We propose a factorisation approach to estimate the meson state matrix elements of the appearing six-quark operators: we assume vacuum saturation for the coupling of the quark fields from the intermediate state, \textit{viz.} we model the meson's non-valence substructure with the vacuum condensate, neglecting the effect of higher excitations. Doing so reduces the operator basis to $Q=\overline  u_\text{L} \gamma^\mu c_\text{L}\otimes \overline  u_\text{L} \gamma_\mu c_\text{L}$ and $Q_\text{S}=\overline  u_\text{R} \, c_\text{R}\otimes \overline  u_\text{R} \, c_\text{R}$. To overcome $\text{SU}(3)_\text{F}$ interference, the effect of the diquark condensate has to be included up to next-to-leading order in the spatial separation between the quark fields, 
\begin{equation}
	\text{
	\raisebox{-8pt}{\parbox[c]{30mm}{
	     \includegraphics[width=30mm]{eqn.eps}
	}
	}}
\,=\,\lvac : \,q\left( x \right) \otimes \overline  q\left( 0 \right)\,:\rvac =  - \frac{{\left\langle {\overline  qq} \right\rangle }}
{{4\nc}}\, \times \,\eins_{\text{c}} \;\left( {\eins_{\text{D}}  - \frac{{{\text{i}}\,m}}
{4}\;\slashed{x}} \right).
\end{equation}
Within this setup, non-vanishing contributions appear in $D=10$ and $D=12$. As expected, we find small $\mathcal{O}(1\%)$ corrections to the $\Gamma_{12}^{ab}$, which reveal a pronounced breaking of $\text{SU}(3)_\text{F}$ symmetry and survive as large effects after the flavour cancellations in (\ref{10081701}). Numerically, 
\begin{equation}
	  \delta\Gamma _{12}  =-(1.8+0.1\,\text{i})\cdot 10^{-5} =  - 0.43\;\lambda _s^2 \;{\overline  z}^{\frac{3}{2}}  + 0.38\;\lambda _s \lambda _b \;{\overline  z}^{\frac{1}{2}}  \;\simeq \; \lambda^{7.3}\;+\;\lambda^{8.2}.
\end{equation}
Note that the contribution to the CKM leading part $\Gamma _{12}^{ss}  - 2\,\Gamma _{12}^{sd}  + \Gamma _{12}^{dd}$ exceeds its $D=6$ value by a factor of 13. The prediction for the mixing rate $y$ likewise is enhanced by a factor $\mathcal{O}(10)$ to $y\lesssim 0.9\cdot 10^{-5} $. The weak phase $\operatorname{arg}\Gamma_{12}$ still remains at the level of $\sim 3\%$. In the light of these results, CP violation at the per cent level does not seem unnatural in the $\langle\overline  q q\rangle$ contribution. 

\vspace{-.5ex}
\section{Future perspectives}
\vspace{-1ex}

Our results are valid in the limit of factorisation and vacuum saturation of the sea-quark contribution. Non-factorisable contributions still need to be quantified. A similar calculation for $M_{12}$ is subject of ongoing work. A result for $M_{12}$ will allow definite predictions of $y$ and the physical weak phase $\phi=\operatorname{arg} (-M_{12}/\Gamma_{12})$. Further efforts could also be directed towards diagram topologies with an intermediate coupling to the  four-quark condensate ($D\geq 12$), from which a second order of flavour violation can be expected, possibly associated with an even larger enhancement of $y$.

\vspace{-2ex}

\vspace{-.5ex}
\begin{thebibliography}{99}
\vspace{-1ex}
\bibitem{exp}
  M.~Staric {\it et al.}  [Belle Collaboration],
  %``Evidence for $D^0$ - $\bar{D}^0$ Mixing,''
  Phys.\ Rev.\ Lett.\  {\bf 98} (2007) 211803
  [hep-ex/0703036];
  %%CITATION = PRLTA,98,211803;%%
	%
  B.~Aubert {\it et al.}  [BABAR Collaboration],
  %``Evidence for $D^0$ -anti-D0 Mixing,''
  Phys.\ Rev.\ Lett.\  {\bf 98} (2007) 211802
  [hep-ex/0703020];
  %%CITATION = PRLTA,98,211802;%%
  %
  T.~Aaltonen {\it et al.}  [CDF Collaboration],
  %``Evidence for $D^0 - \bar{D}^0$ mixing using the CDF II Detector,''
  Phys.\ Rev.\ Lett.\  {\bf 100} (2008) 121802
  [hep-ex/0712.1567].
  %%CITATION = PRLTA,100,121802;%%
  
 %\cite{Barberio:2008fa}
\bibitem{Barberio:2008fa}
  E.~Barberio {\it et al.}  [Heavy Flavor Averaging Group],
  %``Averages of $b-$hadron and $c-$hadron Properties at the End of 2007,''
  hep-ex/0808.1297, and the web-updates at \texttt{http://www.slac.stanford.edu/xorg/hfag/}.
  %%CITATION = ARXIV:0808.1297;%%


%\cite{Georgi:1992as}
%\bibitem{Georgi:1992as}
\bibitem{papersCds}
%\cite{Georgi:1992as}
%\bibitem{Georgi:1992as}
  H.~Georgi,
  %``D - anti-D mixing in heavy quark effective field theory,''
  Phys.\ Lett.\  B {\bf 297} (1992) 353;
  %[hep-ph/9209291].
  %%CITATION = PHLTA,B297,353;%%
%
%\cite{Ohl:1992sr}
%\bibitem{Ohl:1992sr}
  T.~Ohl, G.~Ricciardi and E.~H.~Simmons,
  %``D - anti-D mixing in heavy quark effective field theory: The Sequel,''
  Nucl.\ Phys.\  B {\bf 403} (1993) 605
  [hep-ph/9301212];
  %%CITATION = NUPHA,B403,605;%%
%
%\cite{Bigi:2000wn}
%\bibitem{Bigi:2000wn}
  I.~I.~Y.~Bigi and N.~G.~Uraltsev,
  %``D0 anti-D0 oscillations as a probe of quark-hadron duality,''
  Nucl.\ Phys.\  B {\bf 592} (2001) 92
  [hep-ph/0005089].
  %%CITATION = NUPHA,B592,92;%%
  
\bibitem{inclusive}
%
%\cite{Golowich:2006gq}
%\bibitem{Golowich:2006gq}
  E.~Golowich, S.~Pakvasa and A.~A.~Petrov,
  %``New physics contributions to the lifetime difference in D0 - anti-D0
  %mixing,''
  Phys.\ Rev.\ Lett.\  {\bf 98} (2007) 181801
  [hep-ph/0610039].
  %%CITATION = PRLTA,98,181801;%%


%\cite{exclusive}
\bibitem{exclusive}
  A.~F.~Falk, Y.~Grossman, Z.~Ligeti and A.~A.~Petrov,
  %``SU(3) breaking and D0 - anti-D0 mixing,''
  Phys.\ Rev.\  D {\bf 65} (2002) 054034
  [hep-ph/0110317];
	%
  A.~F.~Falk, Y.~Grossman, Z.~Ligeti, Y.~Nir and A.~A.~Petrov,
  %``The D0 - anti-D0 mass difference from a dispersion relation,''
  Phys.\ Rev.\  D {\bf 69} (2004) 114021
  [hep-ph/0402204].
  %%CITATION = PHRVA,D69,114021;%%
  
%\cite{Bobrowski:2010xg}
\bibitem{Bobrowski:2010xg}
  M.~Bobrowski, A.~Lenz, J.~Riedl and J.~Rohrwild,
  %``How large can the SM contribution to CP violation in $D^0-\overline  D^0$ mixing
  %be?,''
  JHEP {\bf 1003} (2010) 009
  [hep-ph/1002.4794].
%
%\cite{Bobrowski:2009ng}
\bibitem{Bobrowski:2009ng}
  M.~Bobrowski, A.~Lenz, J.~Riedl and J.~Rohrwild,
  %``How Much Space Is Left For A New Family Of Fermions?,''
  Phys.\ Rev.\  D {\bf 79} (2009) 113006
  [hep-ph/0902.4883].
  %%CITATION = PHRVA,D79,113006;%%
%
\end{thebibliography}
\end{document}